\documentclass[aps,epsf,amsfonts,floats,twocolumn,amssymb,amsmath,groupedaddress,showpacs,floatfix,nofootinbib]{revtex4}
\usepackage{graphicx}
\usepackage{epsf}

\usepackage[]{latexsym}
\usepackage{bm}
\usepackage{amsmath}

\newcommand{\be}{\begin{equation}}\newcommand{\ee}{\end{equation}}
\newcommand{\bea}{\begin{eqnarray}}\newcommand{\eea}{\end{eqnarray}}
\newcommand{\brr}{\begin{array}}\newcommand{\err}{\end{array}}
\newcommand{\bit}{\begin{itemize}}\newcommand{\eit}{\end{itemize}}
\newcommand{\ben}{\begin{enumerate}}\newcommand{\een}{\end{enumerate}}

\newcommand{\ba}{\begin{array}}
\newcommand{\ea}{\end{array}}

\def\1{{_{1}}}\def\2{{_{2}}}

\def\noHe0{:\;\!\!\;\!\!:H_e(0):\;\!\!\;\!\!:}
\def\noHm0{:\;\!\!\;\!\!:H_\mu(0):\;\!\!\;\!\!:}

\def\1{{_{1}}}\def\2{{_{2}}}

\begin{document}

\title{Propensity to spending of an average consumer over a brief period}

\author{Roberto De Luca$^1$}

\author{Marco Di Mauro$^1$}

\author{Angelo Falzarano$^2$}

\author{Adele Naddeo$^3$}

\affiliation{ $^1$ Dipartimento di Fisica E.R.Caianiello, Universit\`{a} di Salerno, Fisciano (SA) - 84084, Italy}

\affiliation{ $^2$ Dipartimento di Scienze Economiche e Statistiche, Universit\`{a} di Napoli Federico II, Napoli - 80126, Italy}

\affiliation{ $^3$ INFN Sezione di Napoli, Napoli - 80126, Italy}


\begin{abstract}

Understanding consumption dynamics and its impact on the whole economy and welfare within the present economic crisis is not an easy task. Indeed the level of consumer demand for different goods varies with the prices, consumer incomes and demographic factors. Furthermore crisis may trigger different behaviors which result in distortions and amplification effects. In the present work we propose a simple model to quantitatively describe the time evolution over a brief period of the amount of money an average consumer decides to spend, depending on his/her available budget. A simple hydrodynamical analog of the model is discussed. Finally, perspectives of this work are briefly outlined.

\end{abstract}

\maketitle

\section{Introduction}

Despite their diversity, crises have as common denominator the fact that the consumer is put into a state of alert. As a consequence, a series of behaviors, whose purpose is to limit or completely avoid a given consumption, are activated.

Consumers refer to media and formal and informal communication channels, which sustain the state of alert. Besides spreading news and information, media can have amplifying and distorting effects \cite{Slovic}\cite{Kasperson}, which may have influence on risk perception and assessment. Amplification generally occurs at two stages: the transfer of information about the crisis and the response mechanisms of the society, which in turn may result in secondary impacts on the consumer behavior with rippling effects across time, space and social institutions.

In general, the issue of information diffusion and aggregation in social networks has been widely addressed \cite{Castellano}. Indeed many micro-economic models show that agents may fail at aggregating information, when riding on the information gathered by others without looking for independent sources (for instance public information releases) \cite{herding}. Further studies point out the negative impact of dominant groups on the information aggregation process \cite{golub}, also when agents update their state of knowledge following Bayesian learning schemes \cite{livan}. In the last case a transition has been found to take place, as a function of the noise level in the information available at the beginning, from a regime where information is properly aggregated to one characterized by the failure of such a process \cite{livan}.

Moreover, the consumer, being aware of the non existence of the ``zero risk'',  is oriented in the short term towards particular classes of goods, which guarantee greater safety. Among consumer goods \cite{Nelson}\cite{Andersen}, we distinguish three classes. The first class is given by \emph{convenience goods}, which can be found readily, and do not require the consumer to go through an intensive decision-making process. In the second class there are \emph{shopping goods}, whose cost is higher and which require a comparison based upon their suitability, quality, price, style and so on. The third class comprises \emph{specialty goods}, which require more time and money with respect to shopping goods. Usually the offer of these goods is limited, therefore in this last case the comparison factor is absent and the consumer tends not to be biased, not even by the price. Furthermore, most traded goods have qualities which are difficult or impossible to detect; this lack of information can also affect the consumer's decision-making process \cite{Andersen}.

For these and other reasons, the description of an average consumer behavior in the present economic crisis is not an easy task to accomplish. However, this is a very important problem, especially in view of the links between individual consumption decisions and outcomes for the whole economy (see e.g \cite{DeatonBook}), as the recent attribution of the 2015 Nobel Prize for Economic Sciences to A. Deaton (part of which was indeed motivated by ``the studies of
the link between consumption and income that he conducted around 1990'') clearly shows \cite{NobelPrize}. In fact, very many variables may be needed to describe the propensity to spending of individuals and the relation between income and consumption . Nevertheless, basic conservation laws and ad-hoc assumptions may lead to define the general dynamics of a consumer's available budget.

In this paper we present a simple model to quantitatively describe the time evolution of the amount of money an average consumer decides to spend, depending on his/her available budget. The notion of average consumer, as an instance of the more general concept of \emph{representative agent}, is subject to all the limitations of the latter, in particular it does not take into account the differences between individuals, assuming that they all behave in the same way. The use of the representative agent in economy has been criticized mainly in \cite{Kirman1992}. Progress in providing alternatives to the ad hoc assumptions of the representative agent approach has recently been made by using the tools of Statistical Mechanics (see e.g. \cite{DeMartinoMarsili1} and references therein), and the results have been applied to the specific problem of modeling consumer behavior and preferences \cite{DeMartinoMarsili2, Bardoscia}. It would be interesting to extend the present model in that direction.

An interesting hydrodynamic analog of our model is introduced in order to establish a relation between the consumer's expenditure rate $c(t)$ and the disposable budget $b(t)$. In this way an ordinary differential equation is obtained, whose fixed points analysis shows the behavior of a consumer with a fixed rate of income. Section 2 is devoted to the introduction and solution of the model while in Section 3 some comments and perspectives of this work are outlined.

\section{The model}

In the following we shall denote by $b_t$ the budget the consumer has at his disposal at time $t$. Let us suppose that he/she has an income $y_t$ (e.g. the salary earned in a week) and expenses $c_t$. Moreover, let $r$ be the interest rate. Then the variation of the budget from time $t$ to time $t+1$ (that is, the savings) is given by \cite{DeatonBook,Campbell987}:
\begin{equation}
b_{t+1}-b_t = y_t-c_t+\frac{r}{r+1}b_t,
\end{equation}
where the last term represents the earned interests. This difference equation governs the dynamics of the budget of the consumer. This dynamics can be approximated by a continuous evolution governed by the differential equation:
\begin{equation}\label{dyneq}
\frac{db}{dt}(t)=y(t)-c(t)+\frac{r}{r+1}b(t).
\end{equation}
In the following we shall consider the earned interest to be small compared with all the other quantities in Eq. (\ref{dyneq}). It is then clear that the budget $b(t)$ grows in intervals of time where $y(t)>c(t)$, decreases if $y(t)<c(t)$ and is constant over periods of time in which $y(t)=c(t)$. In the present paper we study the dynamical system defined by Eq. (\ref{dyneq}) under the assumptions that $r \simeq 0$ and that there exists a functional relation between $c$ and $b$, i.e. $c(t)=c(b(t))$. In this way we can give a quantitative account of the time dependence of $b(t)$. We focus only on the single consumer dynamics. The macroeconomic aspects, i.e., the interaction with the economy of the country and in particular the time evolution of the public debt, will be addressed in a forthcoming paper \cite{noi}.

\begin{figure}[htb]
\includegraphics[scale=0.7]{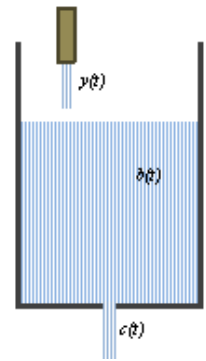}
\caption{A simple hydrodynamic analog to describe consumers' budget dynamics. Here $b(t)$ is the volume of the liquid in the vessel at time $t$ and $y(t)$ and $c(t)$ are the flux flow rates in and out of the bucket.}
\label{Fig1}
\end{figure}

Eq. (\ref{dyneq}) with $r\simeq 0$, reads:
\begin{equation}\label{dyneq2}
\frac{db}{dt}(t)=y(t)-c(t).
\end{equation}
This system has a simple hydrodynamic analog, which is shown in Fig.\ref{Fig1}.

\begin{figure}[htb]
\includegraphics[scale=0.8]{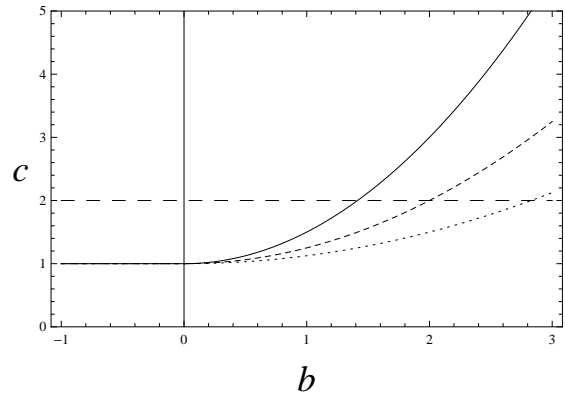}
\caption{Dependence of the function $c(t)$, describing the consumer's expenditure rate, on the disposable budget $b(t)$, for $c_0=1$ and $a=1/8$ (dotted line), $a=1/4$ (dashed line), $a=1/2$ (solid line). The quadratic dependence is justified in the text.}
\label{Fig2}
\end{figure}

It is possible to show, by using Bernoulli's theorem, that the flux rate $c(t)$ depends on $b(t)$ (see e.g. \cite{hydroanalog}). In particular, in this hydrodynamic case, we have that $c(t)\sim \sqrt{b(t)}$. Building on this analogy, we assume that also in our case there is a functional relation between $c$ and $b$, albeit a different one. In particular, we set
\begin{eqnarray}\label{ansatz}
c(t)=\begin{cases}
    a[b(t)]^2+c_0& \text{if } b\geq 0\\
    c_0              & \text{if } b<0,
\end{cases}
\end{eqnarray}
where $a>0$. This relation is depicted in Fig. \ref{Fig2}
and it is supported by the following argument. When the budget is low, say below an arbitrarily defined value $b_F$, the consumer would like to reduce expenses on non strictly necessary goods. However, there is a minimum $c_0$, which represents the minimum expenditure rate allowing an acceptable lifestyle, below which it is not possible to go. On the other hand, if the budget is high, the consumer tends to spend more; in this respect, also the choice to conform to leisure classes plays a significant role \cite{veblen}. We also assume that, if the consumer has debts, that is $b(t)<0$, he just spends the minimum and nothing more.

When we plug (\ref{ansatz}) in the dynamical equation (\ref{dyneq2}) we obtain, in the case $b\geq 0$, the following ordinary differential equation:
\begin{equation}\label{dyneq3}
\frac{db}{dt}+ab^2=y_0-c_0
\end{equation}
where we have set $y(t)=y_0=const.$, so we are describing the behavior of a consumer with a fixed rate of income. We immediately see that
\begin{equation}
\frac{db}{dt}=0\quad\Leftrightarrow\quad (b^{*})^2=\frac{\gamma}{a}
\end{equation}
where we have defined $\gamma=y_0-c_0$. We thus have a fixed point at $b_s=\sqrt{\gamma/a}$, where the positive sign of the root is understood. Clearly we have to set $y_0>c_0$ to ensure an acceptable lifestyle to the consumer at this fixed point. On the other hand, if $y_0<c_0$, there is no fixed point at all. We will study these two cases separately. Before doing so, we observe that the fixed point at $b_s=\sqrt{\gamma/a}$ is stable, as it can be proven by the usual linear stability analysis.

Let us also briefly comment on the case $b<0$. The dynamical equation in this case reads simply
\begin{equation}
\frac{db}{dt}=y_0-c_0
\end{equation}
whose solution is
\begin{equation}
b(t)=\gamma t + b_0
\end{equation}
where, of course, $b_0=b(t=0)$ so that we have a (marginally stable) fixed point in $b=b_0$ if $\gamma=0$.

\begin{figure}[htb]
\includegraphics[scale=0.8]{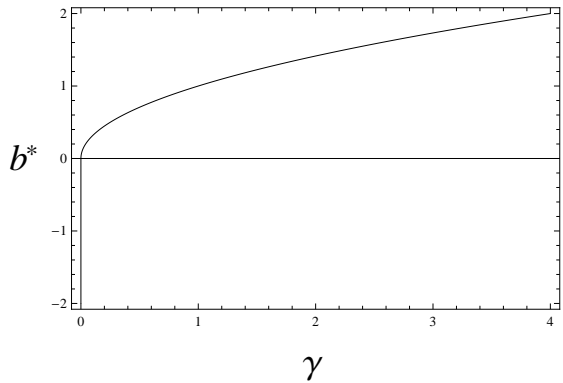}
\caption{Fixed point $b^*$ as a function of $\gamma$.}
\label{Fig3}
\end{figure}

The whole situation is depicted in Fig.\ref{Fig3}. The meaning is obvious, namely that in the equilibrium situation, the higher the income, the higher the budget, even if the expenses grow with the budget. In the case in which there are debts, the only possibility to have equilibrium is to have an income equal to the expenses.

\begin{figure}[htb]
\includegraphics[scale=0.8]{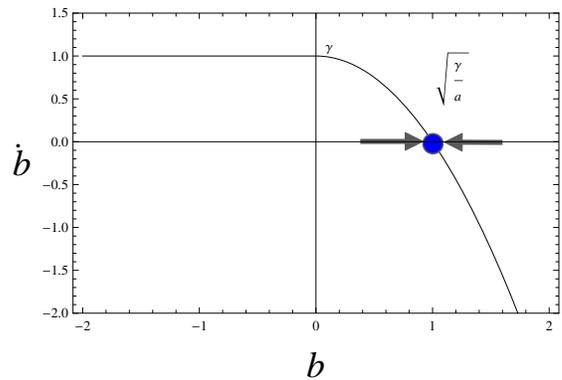}
\caption{Phase space portrait of the dynamical system for $\gamma>0$. Here for definiteness we took $\gamma=1$ and $a=1$. The system shows a stable fixed point at $b=\sqrt{\gamma/a}$}.
\label{Fig4}
\end{figure}

\begin{figure}[htb]
\includegraphics[scale=0.8]{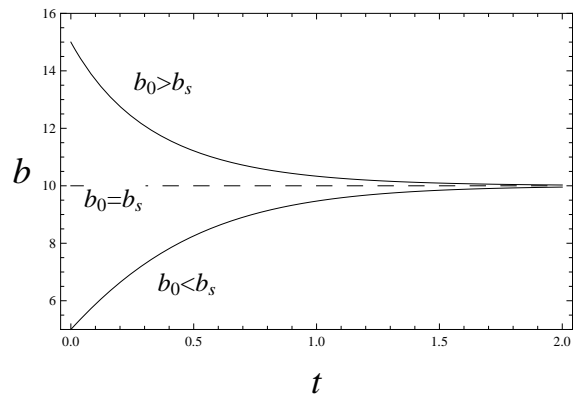}
\caption{Graphical representation of the solution $b(t)$ for $\gamma>0$. Here we took $a=1/8$, $b_s=10$ and $b_0=15$ (upper line), $b_0=b_s=10$ (middle line) and $b_0=5$ (lower line).}
\label{Fig5}
\end{figure}

Let us now focus on the case $b\geq0$. The qualitative behavior of the system for $\gamma>0$ is described by the phase portrait shown in Fig.\ref{Fig4}
The explicit solution of equation (\ref{dyneq3}) in this case is given by:
\begin{equation}\label{Solpositivegamma}
b(t)=b_s\frac{b_0+b_s\tanh(ab_st)}{b_s+b_0\tanh(ab_st)}.
\end{equation}
This function is plotted in Fig.\ref{Fig5} for some sample values of the parameters $b_s$ and $b_0$.
The solution approaches monotonically the asymptotic value $b=b_s$, from above if $b_0>b_s$ and from below if $b_0<b_s$. This means that, when the consumer has no debts, and when the income is higher than the minimal expense $c_0$, he/she tends to spend more or less until a fixed point is reached, that is, until his/her budget stops growing or decreasing, respectively.

\medskip

\begin{figure}[htb]
\includegraphics[scale=0.8]{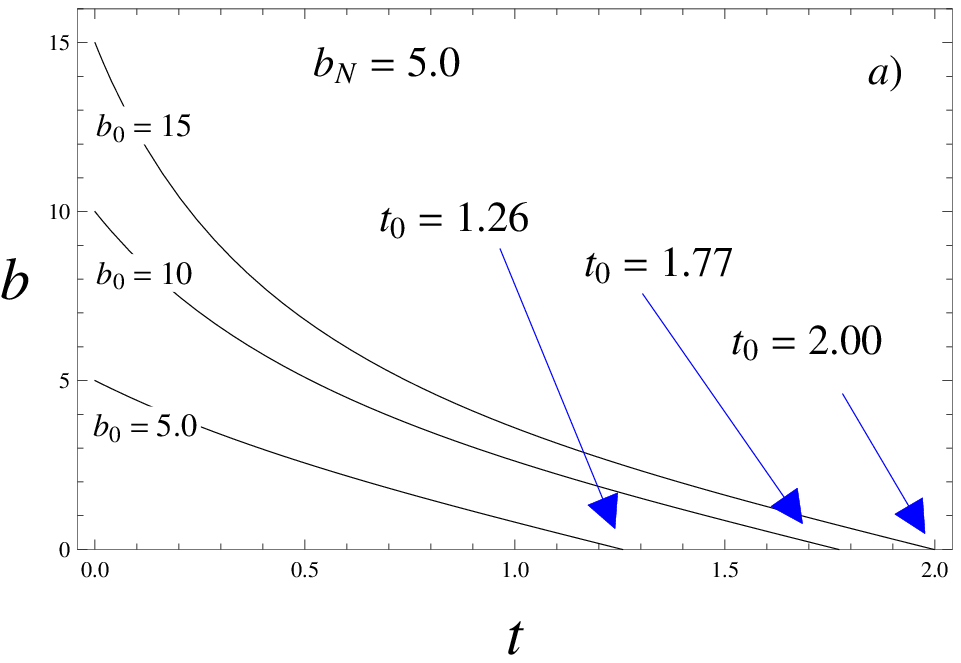}
\includegraphics[scale=0.8]{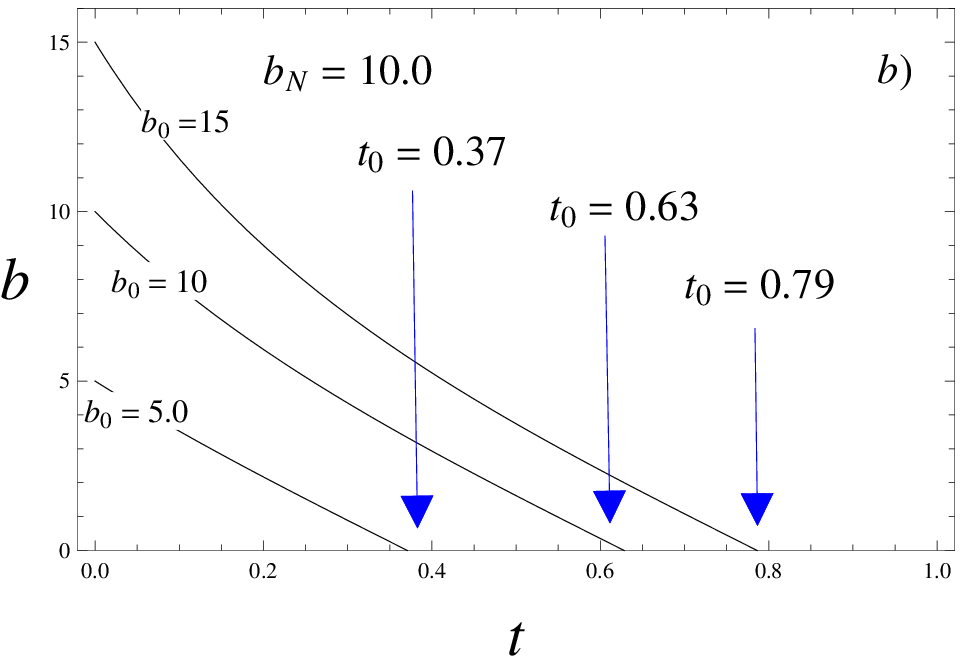}
\caption{Plots of the solution $b(t)$ for $\gamma<0$ with $a=1/8$ and different values of $b_0$ and $b_N$. The corresponding numerical values for the vanishing budget times $t_0$ are reported.}
\label{Fig6}
\end{figure}

Let us now consider the case $\gamma<0$. By defining
\begin{equation}
b_N^2=\frac{|y_0-c_0|}{a}
\end{equation}
we can rewrite Eq.(\ref{dyneq3}) as:
\begin{equation}
\frac{db}{dt}=-(b^2+b_N^2)a
\end{equation}
which can be solved by separation of variables. The solution reads:
\begin{equation}\label{Solnegativegamma}
b(t)=b_N\tan\left[\arctan\frac{b_0}{b_N} - b_N a t \right]
\end{equation}
In this case the budget does not flow to a nonzero fixed point, but it goes to zero. This is obvious since when $\gamma<0$ the minimal expense exceeds the income. It is interesting to predict the time $t_0$ in which the budget goes to zero. By setting $b(t_0)=0$ we have
\begin{equation}
t_0=\frac{1}{b_N a}\arctan\frac{b_0}{b_N}.
\end{equation}
The results for some sample values of the parameters are given in Fig.\ref{Fig6}a-b.

Clearly, our results (\ref{Solpositivegamma}) and (\ref{Solnegativegamma}) can be considered realistic only over the short period, since it is to be expected that the consumer, depending on the circumstances, changes his spending habits over longer periods. In particular it is to be expected that he/she would not allow his budget to monotonically decrease to zero as in eq. (\ref{Solnegativegamma}) by modifying his/her behavior.

\medskip

\section{Discussion and conclusions}

Deaton's work \cite{DeatonBook} points out how a better understanding of the dependence of consumption and saving on the development of income could be gained by studying individual consumptions levels as a function of individual income levels. In this way a closer agreement with macroeconomic data is obtained. Furthermore, knowing the evolution of savings over time in a country is crucial for capital formation and business cycles. So, the main problem is: how much of their incomes do people consume in various time periods? In this paper, starting from the simple assumption of a fixed rate of income, we built up a mathematical model aimed at giving short time quantitative predictions for the behavior of an average consumer within the present economic crisis. A simple hydrodynamic analog has been introduced in order to establish the functional relation between the consumer's expenditure rate $c(t)$ and the disposable budget $b(t)$ and to solve the model. A fixed point analysis has been carried out to clarify the dynamics of the budget of the consumer. Our analysis shows that a fixed point is reached only when the income is higher or equal to the minimal expense. In the case in which such a condition is not met, but still there are no debts, predictions on the time in which such budget goes to zero have also been extracted.

The present model is limited in scope by the use of the concept of average consumer and by the fact that it is valid only on the short period. Besides extending it to address such limitations, it would be interesting to investigate the macroeconomic aspect, i.e. the interaction with the economy of the country and in particular the time evolution of the public debt. This issue will be addressed in a forthcoming paper \cite{noi}.

\section*{Author contribution statement}

All the authors contributed equally to the paper.

\end{document}